\documentclass[a4paper]{jpconf}

\usepackage{amsmath,amssymb}
\usepackage{graphicx}
\usepackage{dcolumn}
\usepackage{bm}
\usepackage[dvips]{color}

\def\rnum#1{\expandafter{%
\romannumeral #1}}
\def\Rnum#1{\uppercase\expandafter{%
\romannumeral #1}}
\newcommand{\bol}[1]{\boldsymbol #1}

\begin{document}
\title{NMR relaxation rate in the field-induced octupolar liquid phase
of spin-$\frac{\bol 1}{\bol 2}$ $\bol J_{\bol 1}$-$\bol J_{\bol 2}$
frustrated chains}

\author{Masahiro Sato${}^1$, Toshiya Hikihara${}^2$ and Tsutomu Momoi${}^3$}

\address{${}^1$Department of Physics and Mathematics, Aoyama Gakuin University,
Sagamihara, Kanagawa 252-5258, Japan\\
${}^2$Department of Physics, Hokkaido University, Sapporo 060-0810, Japan\\
${}^3$Condensed Matter Theory Laboratory, RIKEN, Wako Saitama 351-0198, Japan}

\ead{sato@phys.aoyama.ac.jp}

\begin{abstract}
In the spin-$\frac{1}{2}$ frustrated chain with nearest-neighbor
ferromagnetic exchange $J_1$ and next-nearest-neighbor antiferromagnetic
exchange $J_2$ under magnetic field, magnetic multipolar-liquid
(quadrupolar, octupolar, and hexadecapolar) phases are widely expanded from
the saturation down to a low-field regime. Recently, we have clarified
characteristic temperature and field dependence of the NMR relaxation rate
$1/T_1$ in the quadrupolar phase. In this paper, we examine
those of $1/T_1$ in the octupolar phase combining field theoretical method
with numerical data. The relevance of the results to quasi
one-dimensional $J_1$-$J_2$ magnets such as $\rm PbCuSO_4(OH)_2$,
$\rm Rb_2Cu_2Mo_3O_{12}$ and $\rm Li_2ZrCuO_4$ is shortly discussed.
\end{abstract}

\section{Introduction}
\label{sec:Intro}
The spin-$\frac{1}{2}$ chain with nearest-neighbor exchange coupling $J_1$
and next-nearest-neighbor coupling $J_2$ is one of the representatives of
frustrated spin systems~\cite{Diep}.
If $J_2$ is antiferromagnetic (AF) i.e., $J_2>0$,
the system has magnetic frustration regardless of the sign of $J_1$.
Recently, several quasi one-dimensional (1D) edge-sharing
cuprates~\cite{Masuda2,Enderle,Hagiwara,Hase,Kamieniarz,Baran,Wolter,Drechsler}
such as $\rm LiCu_2O_2$, $\rm LiCuVO_4$, $\rm Rb_2Cu_2Mo_3O_{12}$,
$\rm PbCuSO_4(OH)_2$ and $\rm Li_2ZrCuO_4$, have been investigated
as quantum multiferroic materials or frustrated magnets,
and many of them are believed to be approximately described by a
$J_1$-$J_2$ spin chain with ferromagnetic (FM) $J_1$ and AF
$J_2$~\cite{Furukawa,Sato} at least in the high temperature
regime above their ordering temperatures.

Recent theoretical studies have elucidated effects of external magnetic field
$H$ on the $J_1$-$J_2$ chain~\cite{Hikihara1,Sudan}.
It was found that the field induces three multipolar Tomonaga-Luttinger
(TL) liquid phases,
i.e., quadrupolar, octupolar, and hexadecapolar TL-liquid phases,
in the wide region of $(J_1/J_2,H)$. In these three phases, multipolar operators
$\prod^p_{n=1} S^{-}_{j+n} $ with $p=2$, $3$, and $4$ are respectively shown
to exhibit a quasi long-range order. In addition, in these phases, the
longitudinal spin correlator also decays in an algebraic form, 
while the transverse spin correlator obeys an exponential-decay law.

It is generally difficult to experimentally distinguish magnetic multipolar
phases in spin systems. There is no established method of measuring
multipolar correlations, and static quantities such as
specific heat and susceptibility exhibit no large difference 
between the multipolar and standard TL liquids.
Recently, we have shown~\cite{Sato09} for these multipolar TL liquids that
(i) in contrast to usual TL-liquid phase,
nuclear-magnetic-resonance (NMR) relaxation rate $1/T_1$
of the multipolar liquid phases
decreases with lowering temperature near the saturation, and
(ii) gapless points and peak positions of dynamical structure factors 
are clearly different from those of usual AF spin chains.
Furthermore, in Ref.~\cite{Sato10}
we have numerically evaluated $1/T_1$ of the quadrupolar phase as
a function of temperature and magnetic field, combining field theoretical
approach with numerical data. The obtained result demonstrated
that field dependence of $1/T_1$ is also characteristic:
$1/T_1$ decreases with increasing the field except
for the vicinity of saturation.
These features of dynamical quantities would be helpful to
detect multipolar TL liquids in experiments
for the above quasi-1D magnets.
In this paper, using the same method as Ref.~\cite{Sato10},
we determine the field and temperature dependence of $1/T_1$
in the octupolar phase of spin-$\frac{1}{2}$ $J_1$-$J_2$ chain.
The result is expected to be relevant for quasi-1D $J_1$-$J_2$ magnets
with large FM $|J_1|$. 
For instance, $\rm PbCuSO_4(OH)_2$ might be a possible candidate for 
the realization of octupolar TL liquid~\cite{Wolter2}.

\section{Properties of multipolar phases} 
\label{sec:octupolar}
Here we shortly review the low-energy properties of multipolar
TL-liquid phases of spin-$\frac{1}{2}$ $J_1$-$J_2$ chain~\cite{Hikihara1,Sudan}.
The Hamiltonian of $J_1$-$J_2$ chain under magnetic field $H$
is defined as
\begin{eqnarray}
{\cal H} &=& \sum_{n=1,2}\sum_jJ_n{\bol S}_j\cdot{\bol S}_{j+n}
-\sum_jHS_j^z,
\label{eq:Ham}
\end{eqnarray}
where we focus on the case with $J_1<0$ and $J_2>0$. For the saturated
state in high-field regime, two-, three- and four-magnon bound states
with momentum $\pi$, respectively, become the lowest-energy excitation in the regions
$0<|J_1|/J_2<2.699$ ($0.371<J_2/|J_1|<\infty$), $2.720<|J_1|/J_2<3.514$
($0.285<J_2/|J_1|<0.368$), and $3.514<|J_1|/J_2<3.764$
($0.266<J_2/|J_1|<0.285$). With decreasing field $H$ below the saturation field,
these bound magnons are condensed, and the system can be regarded as a hard-core Bose
gas of bound magnons. We call these gases of two-, three-, and four-magnon
bound states, respectively, the second, third, and fourth order multipolar (quadrupolar,
octupolar, and hexadecapolar) TL liquids~\cite{note1}.
These multipolar phases expand from the saturation down to the low-field regime
and turn into the low-field vector-chiral phase with
$\langle ({\bol S}_j\times{\bol S}_{j+1})^z\rangle\neq 0$.
In the $p$-th order multipolar phase, correlation functions of
both the longitudinal spin operator $S_j^z$ and the $p$-th multipolar
operators $\prod_{n=1}^p S_{j+n}^{\pm}$
(i.e., creation/annihilation operator of bound magnons) can be calculated
from the effective TL-liquid theory for the hard-core Bose gas of multi-magnon
bound states~\cite{Hikihara1}.
They are shown to decay algebraically as follows:
\begin{eqnarray}
\langle S^z_0 S^z_j\rangle &=& M^2 +C_0 \cos\Big[(1-2M)\pi j/p\Big]
|j|^{-2\kappa}+\frac{p^2\kappa}{2\pi^2}|j|^{-2}+\cdots,
\label{eq:correlation_1}\\
\langle \prod_{n=1}^p S^+_{n}\prod_{n=1}^p S^-_{j+n}\rangle
&=& (-1)^j C_1 |j|^{-1/(2\kappa)}+\cdots,
\label{eq:correlation_2}
\end{eqnarray}
where $M=\langle S_j^z\rangle$ is the uniform magnetization per site,
$C_m$ are nonuniversal constants, and $\kappa$ is the TL-liquid parameter.
On the other hand, correlators of $q(<p)$-th multipolar operators including
transverse spin operator $S_j^{\pm}$ obey an exponential-decay form
$\sim e^{-|r|/\xi}$~\cite{Sato09} since any $q$-magnon excitation
requires a finite excitation energy to break $p$-magnon bound states.
The correlation length $\xi$ must be proportional to the inverse of
the binding energy $\Delta$ of magnons.
The absence of the quasi-long-range order in the transverse-spin
correlation plays a key role in the following discussion
of NMR relaxation rate.

\section{NMR relaxation rate}
\label{sec:NMR}
For spin systems in solids, NMR relaxation time $T_1$ is usually
determined through weak hyperfine interactions between
electron and nuclear spins. Therefore, we can obtain information about
electron-spin correlations from measurement of $T_1$.
The formula for $T_1$~\cite{Gia_text,Giamarchi,Giamarchi2} is given by
\begin{eqnarray}
1/T_1 &=& \frac{1}{2}A_{\perp}^2[{\cal S}^{+-}(\omega)+{\cal S}^{-+}(\omega)]
+A^2_{\parallel}{\cal S}^{zz}(\omega),
\label{eq:T_1}
\end{eqnarray}
where $A_{\perp,\parallel}$ are the hyperfine coupling constants,
and ${\cal S}^{\mu\nu}(\omega)$ is the local dynamical structure factor
\begin{eqnarray}
{\cal S}^{\mu\nu}(\omega) &=& \int dt e^{i\omega t}
\langle S_j^\mu(t)S_j^\nu(0)\rangle.
\label{eq:dynamical_st}
\end{eqnarray}
The strength of $A_{\perp,\parallel}$ depends on kinds of nuclei,
crystal structures, and direction of field $H$. The value of $\omega$
is fixed to a nearly resonating frequency of nuclear-spin precession motion,
and it is generally much smaller than the energy scale of electron
spin systems. Therefore, we take the limit $\omega/(k_BT)=\omega\beta\to 0$
in the following discussions.
Utilizing these formulas, we will discuss the temperature and field
dependence of $1/T_1$ in the octupolar liquid phase.

\section{Temperature and field dependence of $\bol 1/\bol T_{\bol 1}$}
\label{sec:calculation_NMR}
From Eqs.~(\ref{eq:correlation_1}), (\ref{eq:T_1}) and
(\ref{eq:dynamical_st}) and
field-theoretical method~\cite{Gia_text,Giamarchi,Giamarchi2},
$1/T_1$ in the multipolar liquid phases is obtained as
\begin{eqnarray}
1/T_1&=&A_\parallel^2 C_0 \frac{2a}{u}\cos(\kappa\pi)B(\kappa,1-2\kappa)
\left(\frac{2\pi a}{\beta u}\right)^{2\kappa-1}\nonumber\\
&&+ A_\parallel^2 \frac{p^2\kappa}{\pi}\left(\frac{a}{u}\right)^2\beta^{-1}
+\cdots.
\label{eq:NMR_boso}
\end{eqnarray}
Here, $\beta=1/(k_BT)$,
$u$ is the excitation velocity, $a$ is lattice spacing,
and $B(x,y)$ is the Beta function.
We note that Eq.\ (\ref{eq:NMR_boso}) is valid
for the temperature regime $k_BT \ll \min(u/a, \Delta)$,
where the low-energy effective theory of the bound-magnon Bose gas
is reliable. If spatial dependence of hyperfine couplings is taken account of
beyond the formula (\ref{eq:T_1}), the prefactors $A_\parallel^2$ in
the first and second term of Eq.~(\ref{eq:NMR_boso}) are generally modified
and become different.

The form of $1/T_1$ in Eq.~(\ref{eq:NMR_boso}) consists of
contributions from the longitudinal spin correlation;
The first and second terms in Eq.~(\ref{eq:NMR_boso}) originate from 
the second and third terms in Eq.~(\ref{eq:correlation_1}), respectively.
Remarkably, both terms are power-law type functions of temperature.
On the other hand, since the transverse spin correlation exhibits a gapped
behavior, its contribution to $1/T_1$ is a thermal-activation form
$\sim e^{-\beta\Delta}$. Therefore, if we consider the temperature
region lower than the binding energy $\Delta/k_B$,
the contribution from transverse spin correlation is negligible.
The disappearance of transverse-spin contribution
is a characteristic feature of multipolar TL liquids
and makes the multipolar liquids distinct from the usual TL liquid
in various quasi-1D AF magnets as discussed below.

Parameters $\kappa$, $u$, and $C_0$ can be evaluated as
a function of field $H$ (or magnetization $M$) and $J_1/J_2$
at zero temperature by using density-matrix renormalization
group method~\cite{Hikihara1,Hikihara2}.
Substituting them into Eq.\ (\ref{eq:NMR_boso}),
we obtain the temperature and field dependence of $1/T_1$.
In particular, the temperature dependence of each term
is determined by the TL-liquid parameter $\kappa$.
For the multipolar phases, $\kappa$ is known to increase
almost monotonically with increasing $H$ (or $M$)
and approach unity at the saturation~\cite{Hikihara1,Sudan}.
We can thereby find that $1/T_1$ decreases with
lowering temperature in the high-field region with $\kappa>1/2$,
while it diverges in the low-field region with $\kappa<1/2$~\cite{Sato09}.
This behavior highly contrasts to the usual TL liquid
such as that of the spin-$\frac{1}{2}$ AF chain,
in which one of the longitudinal- or transverse-spin contributions
diverges at $T \to 0$ and $1/T_1$ is a monotonically increasing
function of temperature in sufficiently low-temperature regime.
Furthermore, for the quadrupolar phase ($p=2$), the field dependence
of $1/T_1$ has also been determined rather accurately~\cite{Sato10}.
It has then been shown that
with increasing $H$, $1/T_1$ decreases for low field regime
and increases in the vicinity of the saturation field
(see Fig. 1 of Ref.~\cite{Sato10}).
This peculiar non-monotonic field dependence of $1/T_1$
is expected to be a common property of all the multipolar TL-liquid
phases in the $J_1$-$J_2$ chain and would be also a hallmark of them.

Let us now examine the temperature and field dependence of
$1/T_1$ in the octupolar TL-liquid phase ($p=3$).
We evaluate $1/T_1$ combining the formula~(\ref{eq:NMR_boso})
with numerical estimates of $\kappa$, $u$, and $C_0$
which are obtained in the same manner
as Refs.~\cite{Hikihara1,Hikihara2,Hikihara3}. The coupling constant
$A_\parallel$ is simply set to unity.
Hereafter, we concentrate on the point of $J_2/|J_1|=0.32$ ($J_1/J_2 = -3.125$),
at which the octupolar phase widely exists
for $0.07 \lesssim M < 1/2$ 
($0.022|J_1| \lesssim H < H_c =0.05|J_1|$)~\cite{Hikihara1}.
We note that in the octupolar phase,
the excitation velocity $u$, i.e., the width of low-energy excitation
band, is rather small compared to those of quadrupolar phase~\cite{Sato10}.
For instance, we evaluate $u/a \simeq 0.06 |J_1|\simeq 0.19J_2$ for $M=0.2$
and $u/a \simeq 0.01|J_1|\simeq 0.03J_2$ for $M=0.4$ at the point of
$J_2/|J_1|=0.32$. This is a natural result because the parameter region
of the octupolar liquid phase is closer to
the ferromagnetic instability point $J_2/|J_1|=0.25$ at zero field. 
The magnon binding energy is also evaluated as 
$\Delta\simeq 0.025 |J_1|\simeq 0.078J_2$ for $M=0.2$ and 
$\Delta\simeq0.041 |J_1|\simeq 0.13J_2$ for $M=0.4$ at $J_2/|J_1|=0.32$~\cite{note2}.
Therefore, when we apply the formula~(\ref{eq:NMR_boso}) at this point,
temperature must be low enough, say, $k_BT < 0.01|J_1|$.
We also note that the numerical estimates of $C_0$ 
in the present calculation are subject to a rather large finite-size
effect and the evaluated $1/T_1$ is less accurate than that of
the quadrupolar liquid in Ref.~\cite{Sato10}.
However, we have checked that the result below is stable and
semi-quantitatively unchanged against such an error of $C_0$. 

\begin{figure}
\includegraphics[width=16pc]{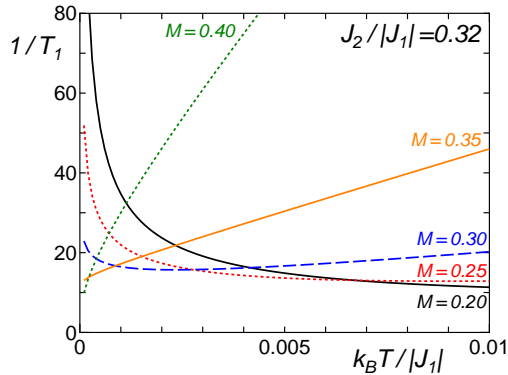}\hspace{2pc}%
\begin{minipage}[b]{14pc}\caption{
Temperature dependence of NMR relaxation rate $1/T_1$
in the octupolar TL-liquid phase for $J_2/|J_1|=0.32$ ($J_1/J_2 = -3.125$).
We simply set the coupling constant $A_\parallel=1$.}
\label{fig:temperature}
\end{minipage}
\end{figure}

\begin{figure}
\begin{minipage}{16pc}
\includegraphics[width=16pc]{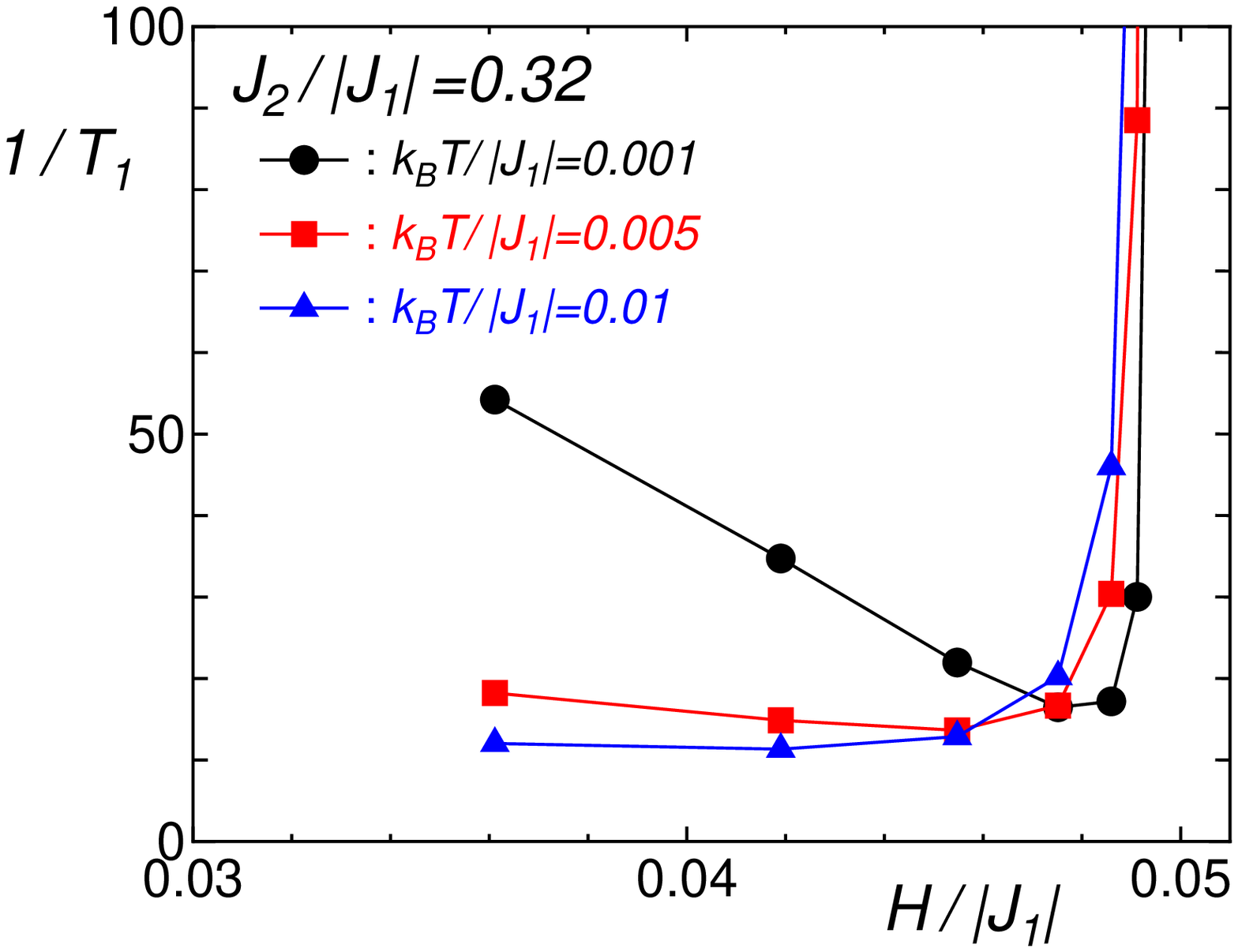}
\caption{
Field dependence of NMR relaxation rate $1/T_1$
in the octupolar TL-liquid phase for $J_2/|J_1|=0.32$ ($J_1/J_2 = -3.125$).
Symbols represent the data obtained from Eq.\ (\ref{eq:NMR_boso})
with numerical values of $\kappa$, $u$, and $C_0$.
Lines are guide for the eyes.
}
\label{fig:field}
\end{minipage}\hspace{2pc}%
\begin{minipage}{16pc}
\includegraphics[width=16pc]{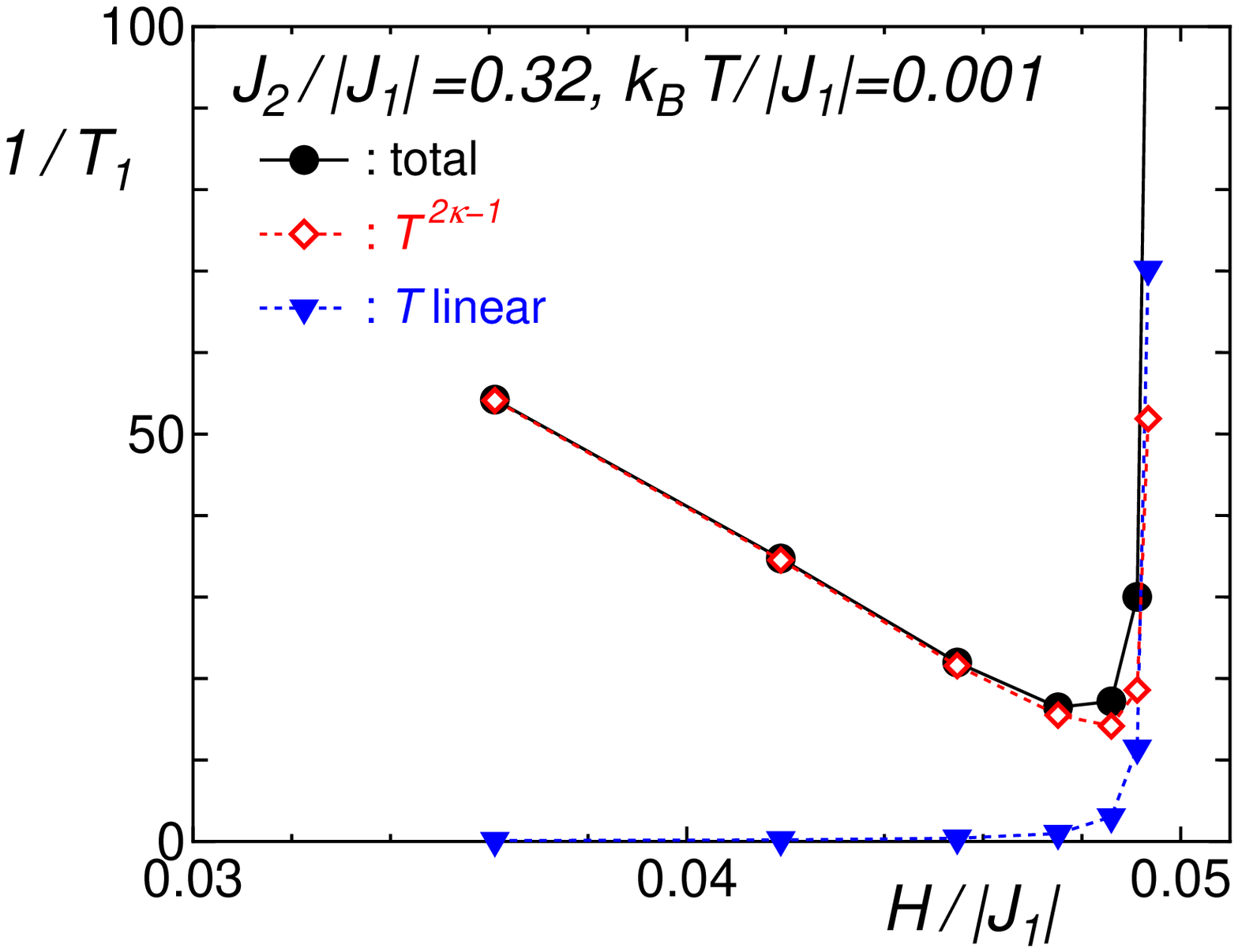}
\caption{
Field dependence of each contribution in NMR relaxation rate $1/T_1$
[Eq.\ (\ref{eq:NMR_boso})]
in the octupolar TL-liquid phase for $J_2/|J_1|=0.32$ ($J_1/J_2 = -3.125$)
and $k_B T/|J_1| = 0.001$.
}
\label{fig:each-comp}
\end{minipage}
\end{figure}

We show in Fig.~\ref{fig:temperature} the temperature dependence of $1/T_1$.
The result confirms the previous prediction that, as temperature decreases,
$1/T_1$ increases divergently for the low-field regime ($\kappa < 1/2$)
and decreases for the high-field regime ($\kappa > 1/2$).
Figure~\ref{fig:field} shows the field dependence of $1/T_1$
for several fixed temperatures.
It is remarkable that, for the sufficiently low temperature
$k_BT=0.001|J_1|$, $1/T_1$ decreases with increasing $H$ (or $M$)
except for the vicinity of the saturation field.
The contribution from each term in Eq.~(\ref{eq:NMR_boso})
is plotted in Fig.~\ref{fig:each-comp} separately.
It is clear that the first term
in Eq.~(\ref{eq:NMR_boso}) proportional to $\beta^{-(2\kappa-1)}$ 
contributes dominantly to a peculiar behavior of
$1/T_1$, decrease with increasing $H$, for a wide field range.
This feature becomes less significant as temperature increases
and the field dependence of $1/T_1$ becomes nearly flat
for a rather high temperature, $k_BT=0.01|J_1|$.
The $T$ and $H$ dependence of $1/T_1$ found above is qualitatively the same as
that of the quadrupolar TL liquid~\cite{Sato10}, as expected.
Even in the neighboring hexadecapolar liquid phase, similar properties must
be realized (although the formula~(\ref{eq:NMR_boso}) would be valid
only in extremely low temperature regime).

Quasi-1D $J_1$-$J_2$ magnets with high crystallographic symmetry often
have principal axes for hyperfine coupling tensor.
For such a case, we can eliminate all the contributions from the
longitudinal spin correlations, i.e., $A_\parallel$ terms, in $1/T_1$,
setting the field $H$ parallel to a principal axis.
As a result, thermal activated contribution $\sim e^{-\beta\Delta}$
proportional to $A_\perp^2$, which is the hallmark of the multipolar liquids,
becomes detectable in $1/T_1$.
In this case, therefore, we can easily distinguish the multipolar TL liquid
and usual TL liquid phases from temperature dependence of $1/T_1$.

\section{Discussions}
\label{sec:diss}
We have studied the NMR relaxation rate $1/T_1$ in
field-induced octupolar TL liquid of the spin-$\frac{1}{2}$
$J_1$-$J_2$ chain.
Substituting numerical estimates of parameters $\kappa$, $u$, and $C_0$
into the field-theoretical result,
we have evaluated $1/T_1$ as a function of temperature and magnetic field.
We have then found that, similarly in the quadrupolar liquid,
$1/T_1$ in the octupolar TL liquid also exhibits
characteristic features which are qualitatively different from
those in the usual TL liquid; In the octupolar liquid,
(i) $1/T_1$ decreases (increases) with increasing temperature $T$
for the low- (high-)field regime, and
(ii) for sufficiently low temperatures, $1/T_1$ decreases
with increasing magnetic field $H$ for a wide field range
except the vicinity of the saturation field.
These features can be understood from the absence of
the quasi-long-range order in the transverse-spin correlation,
which is a natural consequence of the formation of magnon bound states,
and therefore be generally expected in all the multipolar TL liquids in
$J_1$-$J_2$ chain. We note that the above features of $1/T_1$ in
the octupolar phase is observed only in sufficiently low
temperature regime ($k_BT<0.01|J_1|$ at $J_2/|J_1|=0.32$).
This is a quantitative difference from the case of 
the quadrupolar phase, which has relatively small $|J_1|$.

The result in this paper indicates
that NMR relaxation rate $1/T_1$
could be useful to detect a signature of the multipolar TL-liquid
phases. However, only the temperature dependence of $1/T_1$ cannot be a 
sufficient evidence for them.
For instance, $1/T_1$ also decreases with lowering temperature
in high-field spin-gapped states such as magnetization plateaus.
Therefore, we propose that observations of both characteristic behavior 
in $1/T_1$ and gapless behavior of
bulk quantities (specific heat, susceptibilities, etc.) present
an indirect but strong evidence for multipolar TL liquids.

It has been reported~\cite{Hase,Kamieniarz,Baran,Wolter,Drechsler}
that three quasi-1D cuprates $\rm PbCuSO_4(OH)_2$,
$\rm Rb_2Cu_2Mo_3O_{12}$, and $\rm Li_2ZrCuO_4$
have relatively large value of $|J_1|$ ($2\lesssim |J_1|/J_2 \lesssim 4$).
We expect that our result will be helpful for searching for the
appearance of octupolar phase in these compounds under high
magnetic field.

\ack
This work was supported by Grants-in-Aid for Scientific Research
from MEXT, Japan (Grants No.\ 21740277, No.\ 21740295, and No.\ 22014016).



\begin{thebibliography}{9}
\bibitem{note1}
In the narrow range $-2.720<J_1/J_2<-2.669$,
the soft two-magnon bound state has an incommensurate momentum
$k\ne\pi$ to realize a two-component TL liquid. 
In this paper, we do not consider this incommensurate case and focus on
the one-component multipolar TL liquids of the multimagnon bound states
with momentum $k=\pi$. However, the qualitative behavior of $1/T_1$ in the 
incommensurate multipolar phase is expected to be similar to 
those of the one-component multipolar phase since it mainly results 
from the formation of multimagnon bound states.
\bibitem{note2}
We determine the magnon binding energy $\Delta$ in the octupolar TL liquid 
as follows. First we prepare 
$\Delta_n({\cal M})=E_0({\cal M}+n)-E_0({\cal M})$, where 
$E_0({\cal M})$ is the lowest energy eigenvalue in the subspace 
with fixed $\sum_j S_j^z={\cal M}=ML$ 
($L$ is total number of sites). Then, using $\Delta_n$ we define 
four kinds of energy gaps $E_{b1+}=3\Delta_1-\Delta_3$, 
$E_{b2+}=\Delta_1+\Delta_2-\Delta_3$, $E_{b1-}=3\Delta_{-1}-\Delta_{-3}$, 
and $E_{b2-}=\Delta_{-1}+\Delta_{-2}-\Delta_{-3}$. Finally the 
true binding energy is defined as the minimum value of them, 
$\Delta={\rm min}(E_{b1+},E_{b2+},E_{b1-},E_{b2-})$.
\end{thebibliography}
\end{document}